\documentclass[fleqn,twoside]{article}
\usepackage{espcrc2}
\usepackage{graphicx}

\def\be{\begin{equation}}
\def\ee{\end{equation}}
\def\beqn{\begin{eqnarray}}
\def\eeqn{\end{eqnarray}}
\def\no{\nonumber}
\def\CR{\no\\}
\def\ba{\begin{array}{c}}
\def\bat{\begin{array}{cc}}
\def\ea{\end{array}}
\def\bi{\begin{itemize}}
\def\ei{\end{itemize}}

\def\cL{{\cal L}}

\def\cO{{\cal O}}
\def\cR{{\cal R}}
\def\cZ{{\cal Z}}
\def\cY{{\cal Y}}
\def\athdm{A2HDM}
\def\thdm{2HDM}

\newcommand{\e}{\mbox{\rm e}}
\newcommand{\AmS}{{\protect\the\textfont2
  A\kern-.1667em\lower.5ex\hbox{M}\kern-.125emS}}

\title{Flavour constraints on multi-Higgs-doublet models: Yukawa alignment}

\author{A. Pich\address[MCSD]{Departament de F\'{\i}sica Te\`orica, IFIC,
Universitat de Val\` encia - CSIC \\
        Apt. Correus 22085, E-46071 Val\` encia, Spain}
       }

\begin{document}

\begin{abstract}
In multi-Higgs-doublet models, the alignment in flavour space of all Yukawa matrices coupling to a given right-handed fermion guarantees the absence of tree-level flavour-changing neutral couplings, while introducing new sources of CP violation. With $N$ Higgs doublets (and no right-handed neutrinos) the Yukawa Lagrangian is characterized by the fermion masses, the CKM quark mixing matrix and $3(N-1)$ complex couplings.
Quantum corrections break the alignment, generating a minimal-flavour-violation structure with flavour-blind phases. The aligned multi-Higgs-doublet models lead to a rich and viable phenomenology with an interesting hierarchy of flavour-changing neutral current effects,
suppressing them in light-quark systems while allowing potentially relevant signals in heavy-quark transitions.
\vspace{1pc}
\end{abstract}

\maketitle

\section{MULTI-HIGGS-DOUBLET MODELS}

While the $SU(2)_L\otimes U(1)_Y\to U(1)_{\mathrm{em}}$ structure of
spontaneous symmetry breaking is well understood, its
explicit implementation in the electroweak theory is still an open question.
The simplest modification of the Standard Model (SM) Higgs mechanism consists in
incorporating additional scalar doublets, respecting the custodial symmetry,
which can easily satisfy the electroweak precision tests.
This leads to a rich spectrum of neutral and charged scalars,
providing a broad range of dynamical possibilities
with very interesting phenomenological implications.

In their most general version, multi-Higgs-doublet models lead to neutral Yukawa
couplings which are non-diagonal in flavour. These unwanted flavour-changing
neutral current (FCNC) interactions represent a major shortcoming
and have lead to a variety of different implementations, where \emph{ad hoc}
dynamical restrictions are enforced in order to guarantee the suppression of
FCNCs at the experimentally required level.

The most general $N$-Higgs-doublet Yukawa Lagrangian takes the form
\beqn\label{eq:GenYukawa}
\cL_Y \!\!\! & =& \!\!\!\mbox{} -\sum_{a=1}^N\;\left\{ \bar Q_L'\left(
\cY^{(a)'}_d\phi_a\, d_R'
+ \cY^{(a)'}_u\tilde\phi_a\, u_R' \right)
\right.\CR\!\!\! && \hskip .9cm\Bigl.\mbox{}
+ L_L'\, \cY^{(a)'}_l\phi_a\, l_R'\Bigr\} + \mathrm{h.c.}\, ,
\eeqn
where $\phi_a(x)$ are the 
$Y=\frac{1}{2}$ scalar doublets,
$\tilde\phi_a(x) \equiv i \tau_2\,{\phi_a^*}$ their charge-conjugate fields,
$Q_L'$ and $L_L'$ denote the left-handed quark and lepton doublets and $d'_R$,
$u'_R$ and $l'_R$ the corresponding right-handed fermion singlets.
For simplicity, we don't consider right-handed neutrinos, although they
could be easily incorporated.
All fermionic fields are written as $N_G$-dimensional flavour vectors;
the couplings $\cY^{(a)'}_f$ ($f=d,u,l$) are
$N_G\times N_G$ complex matrices in flavour space.

The neutral components of the scalar doublets acquire vacuum expectation values
$\langle 0|\,\phi_a^T(x)|0\rangle = \frac{1}{\sqrt{2}}\, (0\, , v_a\,\e^{i\theta_a})$.
Without loss of generality, we can enforce $\theta_1\! =\! 0$
through an appropriate $U(1)_Y$ transformation.
It is convenient to perform a global $SU(N)$ transformation in the space of scalar fields,
so that only one doublet acquires non-zero vacuum expectation value
$v\equiv\left(v_1^2 + \cdots + v_N^2\right)^{1/2}$. This defines the
so-called Higgs basis, with the doublets parametrized as
\beqn
\Phi_1 \!\!\! & =& \!\!\!\left[ \ba G^+\\ \frac{1}{\sqrt{2}}\left( v + S_1 + i\, G^0\right)\ea \right]\, ,
\CR
\Phi_a \!\!\! & =& \!\!\!\left[ \ba H_a^+\\ \frac{1}{\sqrt{2}}\left( S_a + i P_a\right)\ea \right] \qquad (a=2,\cdots,N)\, .
\no\eeqn
The Goldstone fields $G^\pm(x)$ and $G^0(x)$ get isolated as components of $\Phi_1$. The physical charged (neutral) mass eigenstates are linear combinations of
the fields $H_a^\pm$ ($S_a$ and $P_a$).

In the Higgs basis [$\cY^{(a)'}_f\phi_a = Y^{(a)'}_f\Phi_a$ in (\ref{eq:GenYukawa})],
the fermion masses originate from the $\Phi_1$ couplings,
$M'_f\equiv Y^{(1)'}_f v/\sqrt{2}$.
In general, one cannot diagonalize simultaneously all Yukawa matrices.
Therefore, in the fermion mass-eigenstate basis ($d, u, l, \nu$),
with diagonal mass matrices $M_f$,
the matrices $Y^{(a)}_f$ with $a\not=1$ remain non-diagonal giving rise to FCNC
interactions.

\section{YUKAWA ALIGNMENT}

The unwanted non-diagonal neutral couplings can be eliminated requiring the alignment
in flavour space of the Yukawa matrices \cite{Pich:2009sp}:
%
\beqn\label{eq:alignment}
Y^{(a)'}_{d,l} \!\!\! & =& \!\!\! \varsigma_{d,l}^{(a)}\, Y^{(1)'}_{d,l} = \frac{\sqrt{2}}{v}\, \varsigma_{d,l}^{(a)} M'_{d,l}\, ,
\CR
Y^{(a)'}_{u} \!\!\! & =& \!\!\! \varsigma_{u}^{(a)*}\, Y^{(1)'}_{u} = \frac{\sqrt{2}}{v}\, \varsigma_{u}^{(a)*} M'_{u}\, ,
\eeqn
with $\varsigma_{f}^{(1)}=1$.
In terms of fermion mass eigenstates, $\cL_Y$ takes then the form:
\beqn\label{eq:alignedY}
\cL_Y &\!\!\! =&\!\!\!
-\frac{\sqrt{2}}{v}\;\sum_{a=2}^N\, H^+_a \left\{
\varsigma_{d}^{(a)} \,\bar u_L V M_{d} d_R
\right.\CR &&\hskip .8cm\left.\mbox{}
-\varsigma_{u}^{(a)} \,\bar u_R M_{u}^\dagger V d_L \,
+ \,\varsigma_{l}^{(a)} \,\bar\nu_L  M_l\,l_R\right\}
\CR &&\hskip -1.2cm
-\frac{1}{v}\; \sum_{a=2}^N\, \left[ S_a + i P_a\right]\left\{
\varsigma_{d}^{(a)} \,\bar d_L M_{d} d_R
+ \varsigma_{l}^{(a)} \,\bar l_L M_{l} l_R\right\}\!
\CR &&\hskip -1.2cm
-\frac{1}{v}\; \sum_{a=2}^N\, \left[ S_a - i P_a\right]\,
\varsigma_{u}^{(a)*} \,\bar u_L M_{u} u_R
\CR &&\hskip -1.2cm - \sum_f\; \bar f_L M_f f_R\;
\left\{ 1 + \frac{1}{v}\; S_1\right\}
+\;\mathrm{h.c.}
\eeqn

The flavour alignment results in a very specific structure, with all fermion-scalar
interactions being proportional to the corresponding
fermion masses. The only source of flavour-changing phenomena is
the Cabibbo-Kobayashi-Maskawa (CKM) quark mixing matrix $V$, appearing in the
$W^\pm$ and $H_a^\pm$ interactions. Flavour mixing does not occur in the
lepton sector, because of the absence of right-handed neutrinos.
The Yukawa Lagrangian is fully characterized in terms of the
$3 (N-1)$ complex parameters $\varsigma_{f}^{(a)}$ ($a\not= 1$), which
provide new sources of CP violation without tree-level FCNCs.

\subsection{The aligned two-Higgs-doublet model}

With $N=2$ one obtains the aligned two-Higgs-doublet model
(A2HDM) \cite{Pich:2009sp}, which contains one charged scalar field $H^\pm(x)$
and three neutral mass eigenstates  $\varphi^0_i(x) = \left\{h(x), H(x), A(x)\right\}$,
related through an orthogonal transformation
with the original fields $\mathcal{S}_i = \left\{S_1(x), S_2(x), P_2(x)\right\}$:
$\varphi^0_i(x) = \mathcal{R}_{ij} \mathcal{S}_j(x)$.
The Yukawa Lagrangian is parametrized in terms of the three complex couplings $\varsigma^{(2)}_f\equiv \varsigma_f$, which encode all possible freedom allowed by the alignment conditions. Their flavour-blind phases provide an explicit counter-example to the
widespread assumption that in two-Higgs-doublet models (2HDMs) without tree-level FCNCs all
CP-violating phenomena should originate from the CKM matrix.

In terms of mass eigenstates,
\beqn\label{eq:a2hdm}
\cL_Y^{^{\mathrm{A2HDM}}}\! &\!\!\! =&\!\!\!
- \sum_f\; \bar f_L M_f f_R\;\Bigl\{ 1 + \frac{1}{v}\;\sum_{\varphi^0_i} y_f^{\varphi^0_i}\;\varphi^0_i\Bigr\}
\CR && \hskip -1cm
-\frac{\sqrt{2}}{v}\; H^+ \left\{
\varsigma_{d} \,\bar u_L V M_{d} d_R
-\varsigma_{u} \,\bar u_R M_{u}^\dagger V d_L
\right.\CR &&\hskip .4cm\left.\mbox{}
+ \,\varsigma_{l} \,\bar\nu_L  M_l\,l_R\right\}
\, +\,\mathrm{h.c.}\, ,
\eeqn
with
$y_{d,l}^{\varphi^0_i} = \cR_{i1} + (\cR_{i2} + i\,\cR_{i3})\,\varsigma_{d,l}$
and
$y_u^{\varphi^0_i} = \cR_{i1} + (\cR_{i2} -i\,\cR_{i3}) \,\varsigma_{u}^*$.

\begin{table}[tb]
\caption{CP-conserving ${\cal Z}_2$ models ($\tan{\beta}\equiv v_2/v_1$) \cite{Pich:2009sp}.}
\label{tab:z2models}
\renewcommand{\tabcolsep}{1pc} 
\renewcommand{\arraystretch}{1.2} 
\begin{tabular}{@{}lccc}
\hline
Model & $\varsigma_d$ & $\varsigma_u$ & $\varsigma_l$
\\ \hline
Type I & $\cot{\beta}$ & $\cot{\beta}$ & $\cot{\beta}$
\\
Type II & $-\tan{\beta}$ & $\cot{\beta}$ & $-\tan{\beta}$
\\
Type X & $\cot{\beta}$ & $\cot{\beta}$ & $-\tan{\beta}$
\\
Type Y & $-\tan{\beta}$ & $\cot{\beta}$ & $\cot{\beta}$
\\
Inert & 0 & 0 & 0
\\ \hline
\end{tabular}
\end{table}

FCNCs are usually avoided imposing appropriately chosen discrete $\cZ_2$ symmetries
such that only one scalar doublet couples to a given type of right-handed fermion field
\cite{Glashow:1976nt}. The resulting (CP-conserving) models are recovered for the
particular values of $\varsigma_f$ indicated in Table~\ref{tab:z2models}.

\section{QUANTUM CORRECTIONS}

Higher-order corrections induce a misalignment of the Yukawa matrices,
generating small FCNC effects suppressed by the corresponding loop factors.
The possible flavour-changing interactions are enforced to satisfy
the flavour symmetries of the model and, therefore, are tightly constrained
by the special aligned structure of the starting tree-level Lagrangian
\cite{Pich:2009sp}.
The aligned multi-Higgs-doublet Lagrangian remains invariant under the
following flavour-dependent phase transformations of the fermion mass eigenstates
($f=d,u,l,\nu$, $X=L,R$) \cite{Pich:2009sp,Jung:2010ik}:
\beqn
\lefteqn{f_X^i(x)\;\to\; \e^{i\alpha^{f,X}_i}\; f_X^i(x)
\qquad\quad (\alpha^{\nu,L}_i = \alpha^{l,L}_i)\, ,} &&
\CR
\lefteqn{V_{ij} \;\to\; \e^{i\alpha^{u,L}_i}\, V_{ij}\; \e^{-i\alpha^{d,L}_j}\, ,} &&
\CR
\lefteqn{M_{f,ij}\;\to\; \e^{i\alpha^{f,L}_i}\, M_{f,ij}\; \e^{-i\alpha^{f,R}_j}\, .} &&
\eeqn
Owing to this symmetry, lepton-flavour-violating neutral couplings are identically zero to all orders in perturbation theory, while in the quark sector the CKM mixing matrix remains the only possible source of flavour-changing transitions. The only allowed FCNC local operators have the form
\cite{Pich:2009sp,Jung:2010ik}
\beqn
\cO^{n,m}_u &\!\!\! =&\!\!\!
\bar u_L V (M_d^{\phantom{\dagger}} M_d^\dagger)^n V^\dagger (M_u^{\phantom{\dagger}} M_u^\dagger)^m M_u^{\phantom{\dagger}} u_R\, ,
\CR
\cO^{n,m}_d &\!\!\! =&\!\!\!
\bar d_L V^\dagger (M_u^{\phantom{\dagger}} M_u^\dagger)^n V (M_d^{\phantom{\dagger}} M_d^\dagger)^m M_d^{\phantom{\dagger}} d_R\, ,
\label{eq:FCNCoperators}\eeqn
or similar structures with additional factors of $V$, $V^\dagger$ and quark mass matrices.

Quantum corrections have been analyzed at the one-loop level  within the \athdm.
Using the renormalization-group equations \cite{Cvetic:1998uw,Ferreira:2010xe},
one finds that the only induced FCNC structures are \cite{Jung:2010ik}
\begin{eqnarray}\label{eq:FCNCop}
\lefteqn{\mathcal L_{\mathrm{FCNC}} =  \frac{C(\mu)}{4\pi^2 v^3}\; (1+\varsigma_u^*\varsigma_d^{\phantom{*}})\;\sum_i\, \varphi^0_i(x)} &&
\CR &&\times \; \left\{
(\cR_{i2} + i\,\cR_{i3})\, (\varsigma_d^{\phantom{*}}-\varsigma_u^{\phantom{*}})\;
\cO^{1,0}_d
\right.\\ && \hskip .3cm\left.\mbox{}
-\, (\cR_{i2} - i\,\cR_{i3})\, (\varsigma_d^*-\varsigma_u^*)\;
\cO^{1,0}_{u\phantom{d}} \right\}
\; +\; \mathrm{h.c.}\, ,
\nonumber\end{eqnarray}
with $C(\mu) = C(\mu_0)-\log{(\mu/\mu_0)}$.
These FCNC terms vanish identically when $\varsigma_d =\varsigma_u$ ($\mathcal{Z}_2$ models of type I, X and Inert) or $\varsigma_d =-1/\varsigma_u^*$ (types II and Y), as they should,
since the alignment conditions remain stable under renormalization when they are protected by a $\mathcal{Z}_2$ symmetry \cite{Ferreira:2010xe}.
The leptonic coupling $\varsigma_l$ does not induce any FCNC interaction, independently of its value; the usually adopted $\mathcal{Z}_2$ symmetries are unnecessary in the lepton sector.

In the general case, even if one assumes the alignment to be exact at some high-energy scale $\mu_0$,
a non-zero value for $C(\mu)$ is generated when running to a different scale.
An approximate numerical solution to the renormalization-group equations has been recently obtained \cite{Braeuninger:2010td}, in agreement with (\ref{eq:FCNCop}).
The induced FCNC effects are well below the present experimental bounds.

\subsection{Minimal flavour violation}

The phenomenological success of the SM has triggered the interest on
the so-called Minimal Flavour Violation (MFV) scenarios, where all flavour dynamics and
CP violation is assumed to originate in the CKM matrix
\cite{Chivukula:1987py,Hall:1990ac,Buras:2000dm,D'Ambrosio:2002ex}.
At the quantum level the aligned multi-Higgs-doublet model generates an explicit MFV structure
\cite{Pich:2009sp,Jung:2010ik},
but allowing at the same time for new (flavour-blind) CP-violating phases \cite{Kagan:2009bn}.

For vanishing Yukawa couplings the SM has an $SU(N_G)^5$ symmetry under flavour
transformations of $Q_L$, $u_R$, $d_R$, $L_L$ and $l_R$. The whole Lagrangian can be
made formally invariant under this symmetry if the Yukawa matrices are promoted
to flavour spurions transforming in the appropriate way \cite{D'Ambrosio:2002ex}.
Writing all possible higher-dimension invariant operators in terms of the physical fields
and the Yukawa spurions, one gets then the allowed MFV structures.

MFV within the context of the type II \thdm\ ($\cY^{(2)'}_d= \cY^{(1)'}_u=0$)
was discussed in \cite{D'Ambrosio:2002ex},
in terms of the spurions $\cY^{(1)'}_d$ and $\cY^{(2)'}_u$,
and flavour-blind phases have been recently added to
the resulting structure in \cite{Buras:2010mh}.
These references perform a perturbative expansion around the usual $U(1)_{PQ}$ symmetry limit (type II) and look for $\tan{\beta}$--enhanced effects.
Since the aligned model does not assume any starting ad-hoc symmetry, it leads to a
more general MFV framework with $\tan{\beta}$ substituted
by the dimension $6 (N-1)$ parameter space spanned by the couplings $\varsigma_f^{(a)}$.
While giving rise to a much richer phenomenology, it implies
an interesting hierarchy of FCNC effects, avoiding the stringent experimental constraints for light-quark systems and allowing at the same time for potentially relevant signals in
heavy-quark transitions \cite{Jung:2010ik}. Notice that in the general case, without
$U(1)_{PQ}$ or $\mathcal{Z}_2$ symmetries, $\tan{\beta}$ does not have any physical
meaning because it can be changed at will through $SU(2)$ field redefinitions in the scalar space; the physics needs to be described through the (scalar-basis independent)
parameters $\varsigma_f^{(a)}$.

The spurion formalism is very transparent in the Higgs basis.
Imposing the following transformation properties under $SU(N_G)^5$,
\beqn
Y_d^{(1)}\sim (N_G,1,\bar N_G,1,1)\, ,&&
\CR
Y_u^{(1)}\sim (N_G,\bar N_G, 1,1,1)\, , &&
\CR
Y_l^{(1)}\sim (1,1,1,N_G,\bar N_G)\, , &&
\eeqn
the aligned Yukawa Lagrangian is invariant under the full flavour symmetry.
The operators in Eq.~(\ref{eq:FCNCoperators}) are just the neutral
components of the invariant structures
\beqn
\bar Q'_L \left(Y_d^{(1)'} Y_d^{(1)'\dagger}\right)^n \left(Y_u^{(1)'} Y_u^{(1)'\dagger}\right)^m Y_u^{(1)'} u'_R\, ,
&&\CR
\bar Q'_L \left(Y_u^{(1)'} Y_u^{(1)'\dagger}\right)^n \left(Y_d^{(1)'} Y_d^{(1)'\dagger}\right)^m Y_d^{(1)'} d'_R\, . &&
\eeqn

A similar MFV structure in the leptonic sector \cite{Cirigliano:2005ck} could be
obtained by including non-zero neutrino masses through $\nu_R$ fields or dimension-5
operators \cite{Weinberg:1979sa}.

\section{A2HDM PHENOMENOLOGY}

The built-in flavour symmetries protect very efficiently the aligned model from
unwanted effects, allowing it to easily satisfy the experimental constraints. A thorough
phenomenological analysis of the A2HDM is under way. The charged-current sector has
been studied recently \cite{Jung:2010ik}, focusing on observables where the $H^\pm$
contribution can be expected to be the dominant new-physics effect and theoretical uncertainties
can be controlled.

\begin{figure}[t]
\centering{
\includegraphics[width=58mm]{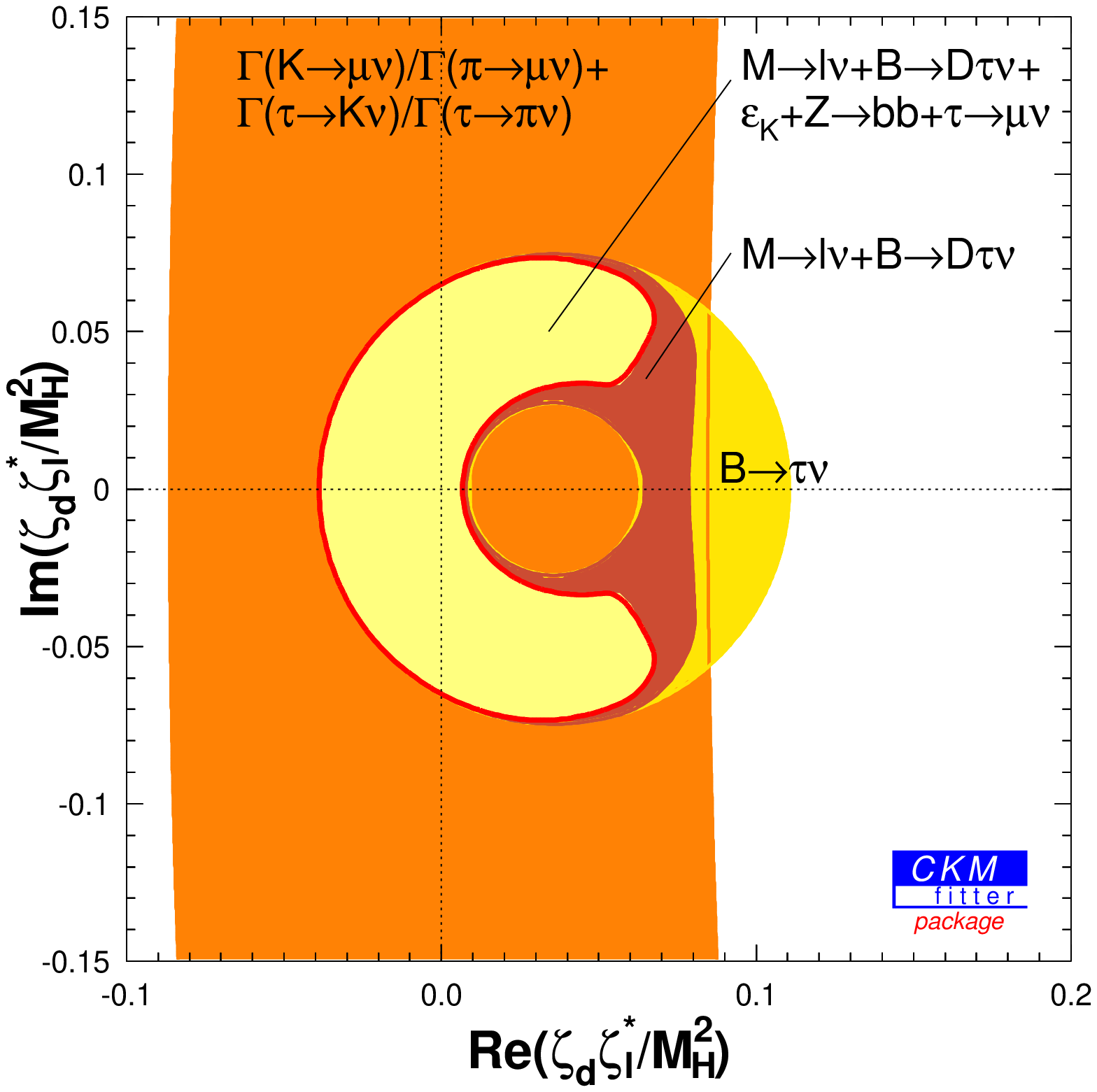}
\vskip .2cm
\includegraphics[width=58mm]{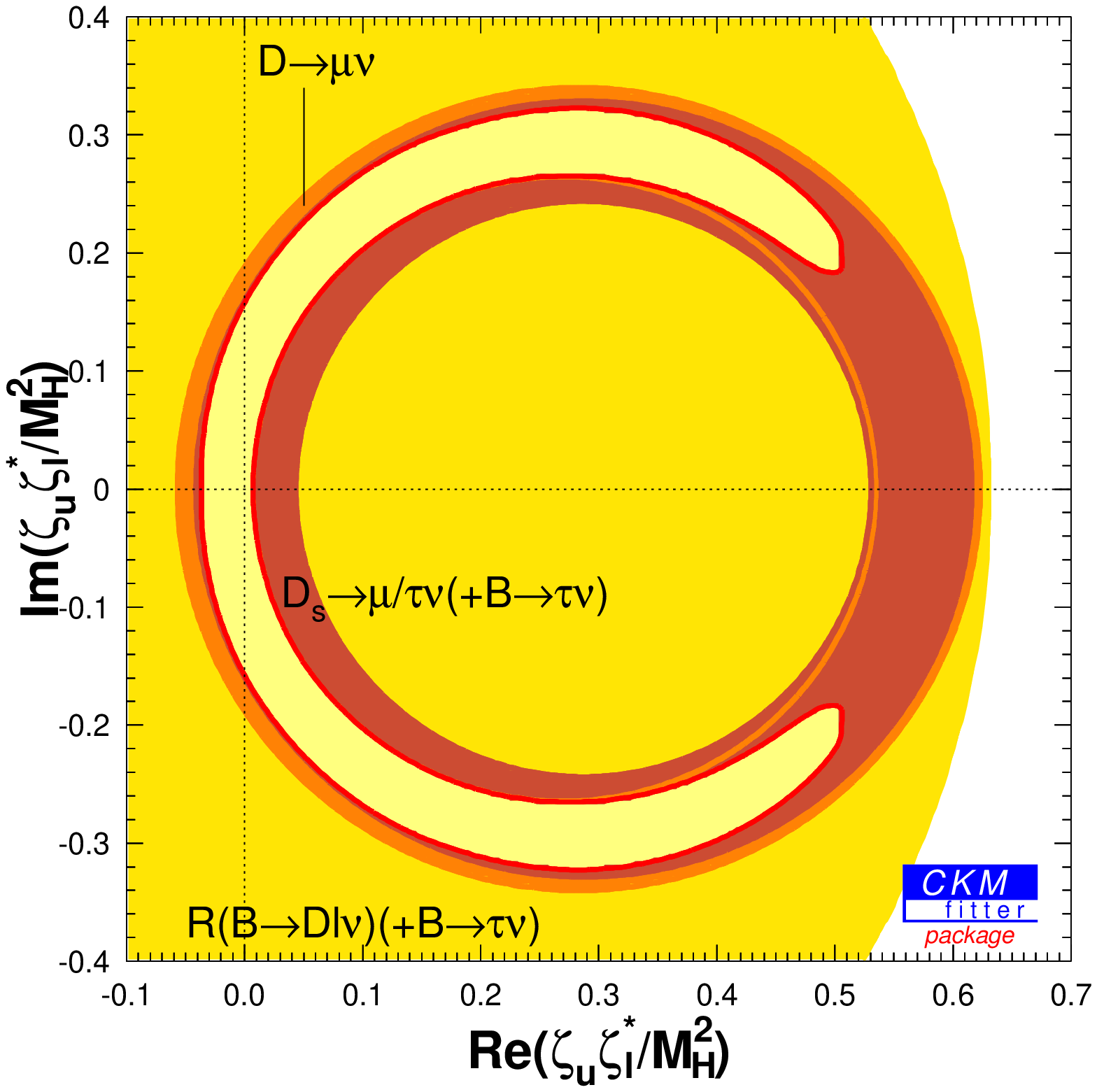}
\vskip -.5cm
\caption{Constraints on $\varsigma_d\varsigma_l^*/M_{H^{\pm}}^2$ (up) and $\varsigma_u\varsigma_l^*/M_{H^{\pm}}^2$ (down) from leptonic and semileptonic decays.
The inner yellow areas show the combined allowed regions at 95\% CL  \cite{Jung:2010ik}.
\label{fig:globalfit}} }
\end{figure}

Leptonic and semileptonic decays are sensitive to tree-level $H^\pm$-exchange contributions,
but the fermion-mass suppression of the Yukawa couplings implies rather weak constraints
on the $\varsigma_f$ parameters. The best direct bound on the leptonic coupling, obtained
from leptonic tau decays, is $|\varsigma_l|/M_{H^\pm}\le 0.40~\mbox{GeV}^{-1}$ (95\% CL).
Semileptonic decays provide information on the products
$\varsigma_{u,d}^*\varsigma_l^{\phantom{*}}/M_{H^{\pm}}^2$; the best limits are obviously obtained
from $B\to\tau\nu$, $D\to\mu\nu$ and $D_s\to (\tau,\mu)\nu$.
The combined constraints from a global analysis of semileptonic processes
are shown in Fig.~\ref{fig:globalfit}.

\begin{figure}[tbh]
\centering{
\includegraphics[width=60mm]{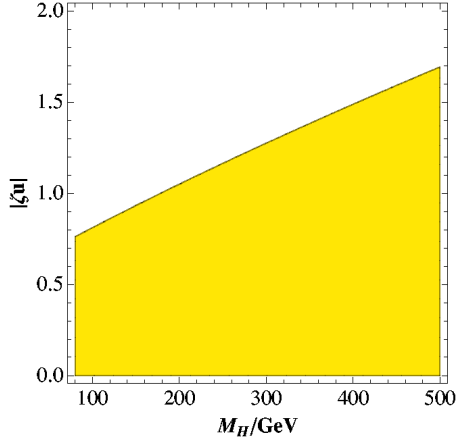}
\caption{95\% CL constraints from $\epsilon_K$ \cite{Jung:2010ik}. \label{fig:EpsK}}}
\end{figure}

More stringent bounds are obtained from loop-induced transitions involving
virtual top-quark contributions, where the $H^\pm$ corrections are enhanced by the
top mass. Direct limits on $|\varsigma_u|$ can be derived from $Z\to b\bar b$,
$B^0$-$\bar B^0$ mixing and the CP-violating parameter $\epsilon_K$.
The last observable provides the strongest
limits, which are shown in Fig.~\ref{fig:EpsK}.
Together with the tau-decay constraint on $|\varsigma_l| /M_{H^\pm}$, this gives the limit
$|\varsigma_u^{\phantom{*}}\varsigma_l^*|/M_{H^\pm}^2<0.005~\mathrm{GeV}^{-2}$, which is much stronger than the information extracted from the global fit to leptonic and semileptonic decays.

\begin{figure}[tb]
\centering{
\includegraphics[width=7cm]{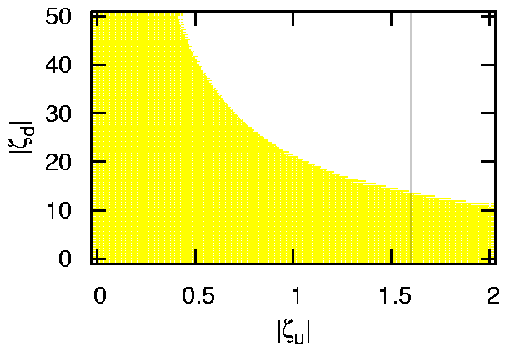}
\vskip .2cm
\includegraphics[width=7.1cm]{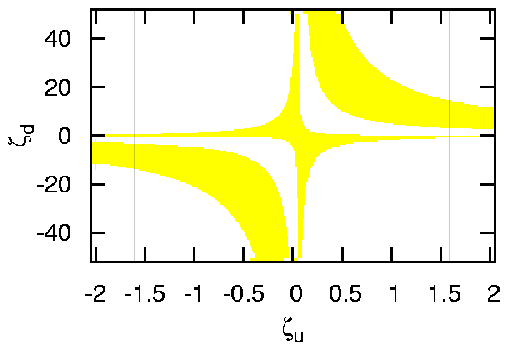}
\caption{\label{fig:ud}
$\bar{B} \rightarrow X_s \gamma$ limits on $\varsigma_{u,d}$ ($95\%$ CL) \cite{Jung:2010ik},
for complex (upper plot) and real (lower plot) couplings.
The black lines indicate the upper limit on $|\varsigma_u|$ from $\epsilon_K$.
}}
\end{figure}

The radiative decay $\bar B\to X_s\gamma$ provides another important source of information.
The $H^\pm$ contributions modify the Wilson coefficients of the low-energy SM
operators with two terms of similar size:
$C_i^{\mathrm{eff}}=C_{i,SM}+|\varsigma_u|^2\, C_{i,uu}-(\varsigma_u^*\varsigma_d^{\phantom{*}})\, C_{i,ud}$.
Their combined effect can be quite different depending on the value of
the relative phase $\varphi\equiv\arg{(\varsigma_u^*\varsigma_d^{\protect\phantom{*}})}$.
This results in rather weak limits on $|\varsigma_{u,d}|$ because a destructive interference can be adjusted through the phase $\varphi$.
Scanning $\varphi$ in the whole range from $0$ to $2\pi$ and taking
$M_{H^\pm}\in[80,500]$~GeV, one obtains the
constraints shown in Fig.~\ref{fig:ud} (upper plot). One finds roughly $|\varsigma_d| |\varsigma_u|<20$ (95\% CL).
Much stronger bounds are obtained at fixed values of $\varphi$. This is shown in the lower plot,
where $\varsigma_u$ and $\varsigma_d$ have been assumed to be real
(i.e. $\varphi=0$ or $\pi$);
couplings of different sign are then excluded, except at very small values, while a broad region of large equal-sign couplings is allowed, reflecting again the possibility of a destructive interference.

In the type II and Y \thdm s, where  $\varsigma_u \varsigma_d = -1$,
the decay $\bar{B}\to X_s\gamma$ provides a very strong lower bound on the charged scalar mass, $M_{H^\pm}> 277~\mathrm{GeV}$ (95\% CL), due to the constructive interference of the two contributing amplitudes \cite{Jung:2010ik}. This bound disappears in the more general \athdm,
but a strong correlation among the allowed ranges for $M_{H^\pm}$ and $\varsigma_{u,d}$ remains.

Another important observable is the CP rate asymmetry,
\begin{equation}
a_{CP}= \frac{BR(\bar{B}\to X_{s}\gamma)-BR(B\to X_{\bar{s}}\gamma)}{BR(\bar{B}\to X_s\gamma)+BR(B\to X_{\bar{s}}\gamma)} \,,
\end{equation}
which is very small in the SM.
Once the constraints from the branching ratio are implemented, the A2HDM predicts an asymmetry smaller than the present experimental bounds.
A sizable Yukawa phase $\varphi$ could generate values of $a_{CP}$ large enough to be relevant for future high-precision experiments. However, a NNLO analysis of the theoretical prediction appears to be needed to reduce the presently large theoretical uncertainties and fully exploit such a measurement \cite{Jung:2010ik}.

While including as limiting cases all known $\cZ_2$ models, the \athdm\ provides new sources
of CP violation through the $\varsigma_f$ phases.
Their compatibility with all presently measured observables and the possibility of generating
sizeable CP-violation signals, within the reach of future experiments, are obviously important
questions that need to be investigated. Since Yukawa couplings are proportional to the corresponding
fermion masses, FCNCs and CP-violation effects are suppressed in light-quark systems while potentially relevant signals could show up
in heavy-quark transitions. This tree-level pattern is reproduced in the FCNC operators
(\ref{eq:FCNCoperators}), generated at one loop. Owing to their CKM and mass-matrix factors, the
most relevant terms are the $\bar s\, b$ and $\bar c\, t$ operators. The top quark and
the $B_s^0$-$\bar B_s^0$ meson system appear then to be promising candidates in
the search for interesting effects. For instance, a sizeable $B_s^0$ mixing phase could
be generated either through $H^\pm$ contributions to the mixing amplitude or through
neutral $\varphi^0_i(x)$ exchanges involving the FCNC one-loop operators in
(\ref{eq:FCNCop}) \cite{Jung:2010ik}.

The presence of flavour-blind phases could induce electric dipole moments (EDMs) at a
measurable level \cite{Buras:2010zm}. Direct one-loop contributions to the light-quark EDMs are
strongly suppressed by the light quark masses and/or CKM factors. However, this suppression
is no longer present in some two-loop contributions to the T-odd 3-gluon operator
$\tilde G_{\mu\nu} G^{\mu\alpha} G^\nu_\alpha$, induced by scalar exchanges within a heavy-quark
loop \cite{Weinberg:1989dx,Dicus:1989va}. For values of Im($\varsigma_u^*\varsigma_d^{\phantom{*}}$) of $O(1)$, the predicted neutron EDM could be close to the present experimental upper bound
and within reach of future high-precision measurements \cite{Trott:2010iz}. A detailed
analysis of EDMs within the \athdm\ is in progress.

\section*{ACKNOWLEDGEMENTS}
I would like to thank Martin Jung and Paula Tuz\'on for a very enjoyable collaboration
and the organizers of this workshop for creating such a friendly and stimulating atmosphere.
This work has been supported in part by the EU MRTN network FLAVIAnet [Contract No. MRTN-CT-2006-035482], by MICINN, Spain
[Grants FPA2007-60323 and Consolider-Ingenio 2010 Program CSD2007-00042 --CPAN--] and by Generalitat Valenciana [Prometeo/2008/069].

\end{document}